\documentclass[english,twocolumn,showpacs,preprintnumbers,amsmath,amssymb]{revtex4}

\usepackage{graphicx}
\usepackage{babel}

\begin{document}

\title{Prediction of the collapse point of overloaded materials by monitoring energy emissions}

\author{Srutarshi Pradhan}
\email{srutarshi.pradhan@sintef.no}
\affiliation{SINTEF Petroleum Research, NO-7465 Trondheim, Norway}

\author{Per C. Hemmer}
\email{per.hemmer@ntnu.no}
\affiliation{Department of Physics, Norwegian University of Science and
Technology, N0-7491 Trondheim, Norway}

\begin{abstract}
A bundle of many fibers with stochastically distributed breaking thresholds 
is considered as a model of composite materials. The 
fibers are assumed to share the load equally, and to obey Hookean elasticity 
 up to the breaking point. The bundle is slightly overloaded, which leads to 
complete failure.  We study the properties of emission bursts in which an 
amount of energy $E$ is released. The analysis shows that the size of the 
energy bursts has a minimum when the system is half-way from the collapse point.
\end{abstract}

\pacs{02.50.-r}

\maketitle

\section{Introduction}

During the failure process in composite materials under external load, 
bursts (avalanches) of different magnitudes are produced, where a burst
consists of simultaneous rupture of several elements. 
At each failure, the sudden internal stress redistribution in the material 
is accompanied by a 
rapid release of mechanical energy. Therefore, with each {\em burst} there will 
be a corresponding {\em energy emission} burst. A useful experimental technique 
to monitor 
\footnotemark[0]
such energy bursts is to measure the acoustic emissions, the 
elastically radiated waves produced in the bursts \cite{AE1,AE2}.
\footnotetext[0]{The Scottish Forth Road Bridge is supported by two main 
cables, each with 11618 fibers (wires), some of which have failed. Acoustic
 monitoring was installed in 2006 to detect further snapping of the wires.
(See Forth Road Bridge in Wikipedia.)}

Fiber bundle models, with statistically distributed thresholds for the 
breakdown of the individual fibers, are interesting models of failure 
processes in materials. They are characterised by clear-cut
rules for how stress caused by a failed element is redistributed on undamaged 
fibers. These models have been much studied since they can be analysed to an 
extent that is not possible for more complex materials (For reviews, see 
\cite{Pradhan1,Herrmann,Chakrabarti,Sornette,Sahimi,Bhatta}). The statistical 
distribution of the {\em magnitude} of avalanches in fiber bundles is well 
studied \cite{Hemmer1,Pradhan2,Hemmer2}, and the failure dynamics under 
constant load has been formulated through recursion relations which in turn 
explore the phase transitions and associated critical behavior in these models
 \cite{Pradhan3}.

In this article we show that the catastrophic collapse point of an overloaded 
bundle can be predicted by monitoring the energy emission rate.

We consider a bundle consisting of a large number $N$ of fibers, clamped at 
both ends. We study equal-load-sharing models, in which the load previously 
carried by a failed fiber is shared equally by all the remaining undamaged 
fibers \cite{Pierce,Daniels,Smith,Phoenix}. The fibers obey Hooke's law, such 
that the energy stored at elongation $x$ equals $x^2/2$, where we for
 simplicity have set the elasticity constant equal to unity. Each fiber $i$ is 
associated with a breakdown threshold $x_i$ for its elongation. When its length 
exceeds $x_i$, the fiber breaks immediately, and does not contribute to the 
bundle strength thereafter. The individual thresholds $x_i$ are assumed to be 
independent random variables with the same cumulative distribution function 
$P(x)$ and a corresponding density function $p(x)$.
If an external load $F$ is applied to a fiber bundle, the resulting breakdown 
events can be seen as a sequential process \cite{Pradhan2}. In the 
first step all fibers that cannot withstand the applied stress ($F/N$) break. 
Then the stress is 
redistributed on the surviving fibers, which compels further fibers to fail, 
etc. This iterative process continues until all fibers fail, or an equilibrium 
situation with a nonzero bundle strength is reached. Since the number of 
fibers are finite, the total number of steps, $t_f$, in this sequential 
process is finite.
At the stress (or elongation) $\sigma$ per surviving fiber the total force on 
the 
bundle is $\sigma$ times the number of intact fibers. The expected, or average, 
force at this stage is therefore
\begin{equation}
F(\sigma) = N\;\sigma\;[1-P(\sigma)].
\label{2}\end{equation}
The maximum $F_c$ of $F(\sigma)$ corresponds to the value $\sigma_c$ for which
 $dF/d\sigma$ vanishes. Thus
\begin{equation}
1-P(\sigma_c)-\sigma_c\;p(\sigma_c)=0,
\label{3}\end{equation}
where the critical stress $\sigma_c$ is defined as
\begin{equation}
\sigma_c=F_c/N.
\label{4}\end{equation}
When the applied load is more than $F_c$ (or $\sigma > \sigma_c$),
we call that the bundle is {\em overloaded}.

We can study the stepwise failure process in the fiber bundle when a fixed 
external load $F$ is applied. Then the initial external stress is 
$\sigma=F/N$.  Let $N_t$ be the number of undamaged 
fibers at 
step no.\ $t$, with $N_0=N$. We want to determine how $N_t$ decreases until 
the failure process stops. With $N_t$ intact fibers, an expected number
\begin{equation}
[N\;P(N\sigma/N_t)]
\end{equation}
of fibers will have thresholds that cannot withstand the load, and consequently these fibers break at once. Here $[X]$ denotes the largest integer not exceeding $X$. The number of intact fibers in the next step is therefore
\begin{equation}
N_{t+1}=N-[N\;P(N\sigma/N_t)],
\end{equation}
or, for all practical purposes,
\begin{equation}
n_{t+1}=1-P(\sigma/n_t).
\label{7}\end{equation}
Here $n_t$ denotes the fraction of undamaged fibers at step $t$,
\begin{equation}
n_t=N_t/N
\end{equation}
In each burst a certain amount of elastic energy is released \cite{Pradhan5}, 
and we consider now the energy emission process.  
\section{ENERGY RELEASE MINIMUM}
For a burst in which the number of intact fibers is reduced from $N_{t-1}$ to $N_t$, all fibers with 
thresholds $x$ between the values $F/N_{t-1}=\sigma/n_{t-1}$ and $F/N_t=\sigma/n_t$ break. Since there are $Np(x)\;dx$ fibers with thresholds in $(x,x+dx)$, 
the energy emitted in this burst is
given by
\begin{equation}
E_t = \int_{\sigma/n_{t-1}}^{\sigma/n_t} {\textstyle \frac{1}{2}}x^2\;Np(x)\;dx.
\label{Et}\end{equation}

We consider external stresses that are slightly above the critical value,
\begin{equation}
\sigma = \sigma_c +\epsilon,
\end{equation}
where $\epsilon$ is small and positive. Large stresses are less interesting, 
since the system will then break down quickly (See the Appendix). 
Simulations for a 
uniform threshold distribution, $p(x)=1$ for $0\leq x\leq 1$, show that the 
energy emission has a minimum at some value $t_E(\epsilon)$, and that for 
varying loads the minima all occur at value close to ${\textstyle \frac{1}{2}}$
 when plotted as function of the scaled variable $t/t_f$ (Fig.\ 1). Here $t_f$ 
is the number of iterations corresponding to complete bundle failure. We will 
'show this analytically, and in addition demonstrate that the result is not 
limited to the uniform distribution.  
\begin{figure}
\includegraphics[width=6cm,height=6cm]{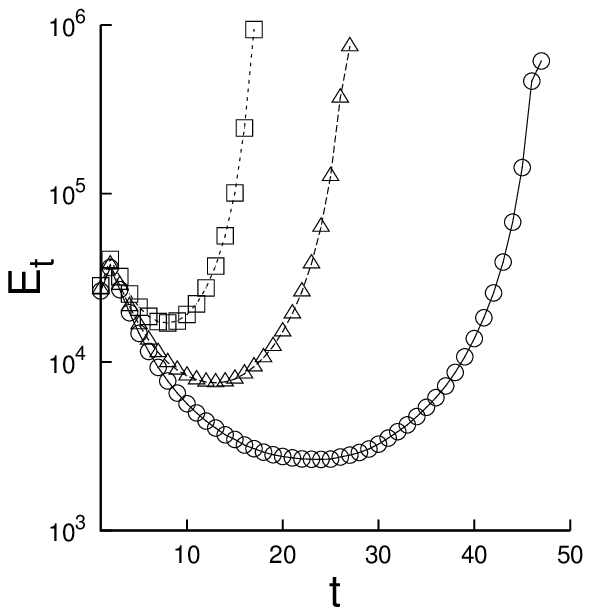}
\includegraphics[width=6cm,height=6cm]{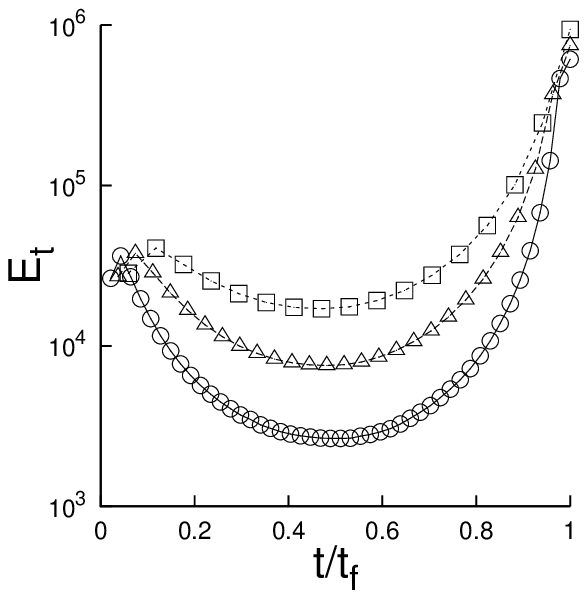}
\caption{Energy emission $E_t$ vs. step $t$ (upper plot)
and vs.\ the rescaled step variable $t/t_f$ (lower plot) for the uniform
threshold distribution for a bundle of $N= 10^7$ fibers.
Different symbols are used for different excess stress levels
$\sigma -\sigma _c$: 0.001  (circles), 0.003 (triangles), 0.007 (squares). }
\label{En-uni}
\end{figure}
Fig. \ref{En-uni}  shows that for large $N$ and near the minimum the energy 
emission $E_t$ appears to be an almost continuous function of $t$. The minimum 
is therefore located where the derivative of $E_t$ with respect to $t$ vanishes. From (\ref{Et}) we obtain
\begin{equation}
dE_t/dt = {\textstyle \frac{1}{2}} N \left[-\sigma^2 n_t^{-4} p(\sigma/n_t)\;\dot{n}_t+
\sigma^2 n_{t-1}^{-4}\;p(\sigma/n_{t-1})\;\dot{n}_{t-1},\right]
\label{dE}\end{equation}
where the dot denotes differentiation with respect to time.

To eliminate quantities at time step $t-1$ we use the connection (\ref{7}), which gives
\begin{equation}
\dot{n}_t = \dot{n}_{t-1}\; p(\sigma/n_{t-1})\;\sigma n_{t-1}^{-2}. 
\end{equation} 
Insertion into (\ref{dE}) yields
\begin{equation}
\frac{dE}{dt} = \dot{n}_t \;\frac{N}{2\sigma}\left[-\frac{\sigma^3p(\sigma/n_t)}{n_t^4}+\frac{\sigma^2}{n_{t-1}^2}\right].
\end{equation}
We also need $n_{t-1}$. Eq.\ (\ref{7}) gives
\begin{equation}
\sigma/n_{t-1} = P^{-1}(1-n_t),
\end{equation}
where $P^{-1}$ denotes the inverse function to $P$. In conclusion,
\begin{equation}
\frac{dE}{dt} = \dot{n}_t\;\frac{N}{2\sigma}\left[-\frac{\sigma^3p(\sigma/n_t)}{n_t^4}+\left(P^{-1}(1-n_t)\right)^2\right].
\end{equation}
Since $\dot{n}_t$ is always negative, the minimum of energy emission occurs 
when the relative number of intact fibers satisfies
\begin{equation}
P^{-1}(1-n_t) = \sqrt{\sigma^3\; p(\sigma/n_t)\;n_t^{-4}},
\end{equation} 
i.e.
\begin{equation}
1-n_t = P\left(\sqrt{\sigma^3\; p(\sigma/n_t)\;n_t^{-4}}\right).
\label{min}\end{equation}
Now we turn to specific cases.\\

\section{EXPLICIT RESULTS}
We start with the simplest case, the uniform distribution, on which the 
simulations in Fig.\ 1 were based.\\

\subsection{Uniform distribution}
For the uniform threshold distribution, $P(x)=x$ for $0\leq x\leq 1$, the 
condition (\ref{min}) takes the form
\begin{equation}
1-n_t=\sqrt{\sigma^3 \;n_t^{-4}}, 
\end{equation}
or
\begin{equation}
(1-n_t)n_t^2 = \sigma^{3/2}.
\label{nuni} \end{equation}
For this case the maximum of the force, Eq.\ (\ref{2}), is $F=N/4$, 
corresponding to the critical stress $\sigma_c = \frac{1}{4}$. For a small 
excess stress $\epsilon = \sigma-\sigma_c=\sigma - \frac{1}{4}$, 
Eq.\ (\ref{nuni}) takes the form
\begin{equation}
(1-n_t)n_t^2 =\left({\textstyle \frac{1}{4}}+\epsilon\right)^{3/2} = {\textstyle \frac{1}{8}}\left( 1+ 6\epsilon + {\cal O}\left(\epsilon^2\right)\right).
\label{euni}\end{equation}
To lowest order $n_t=\frac{1}{2}$, and one shows easily that
\begin{equation}
n_t = {\textstyle \frac{1}{2}} + 3\epsilon
\label{unin}\end{equation}
satisfies (\ref{euni}) to first order in $\epsilon$.

To find $t_E$, the value of $t$ corresponding to (\ref{unin}) we use the 
previously derived solution of the iteration (\ref{7})
\begin{equation}
n_t = {\textstyle \frac{1}{2}} -\sqrt{\epsilon}\;\tan(At-B),
\label{soluni}\end{equation}
where $A=\tan^{-1}(2\sqrt{\epsilon})\simeq 2\sqrt{\epsilon}$ and $B=\tan^{-1}(1/2\sqrt{\epsilon})$.

We see that (\ref{unin}) requires $\tan(At_E-B) = -3\sqrt{\epsilon}$, i.e.\ $t_E-B/A =-3\sqrt{\epsilon}/A $.
To lowest order we therefore obtain
\begin{equation}
t_E = B/A - {\textstyle \frac{3}{2}} .
\label{unitE}\end{equation}

Moreover we see from (\ref{soluni}) that at for 
\begin{equation}
t=t_f=2B/A
\label{unif}\end{equation}
 we have
\begin{equation}
n_{t_f} = {\textstyle \frac{1}{2}} - \sqrt{\epsilon}\;\tan(B) =0,
\end{equation}
 signifying complete collapse of the fiber bundle. By comparison between (\ref{unitE}) and (\ref{unif}) we obtain
\begin{equation}
t_E = {\textstyle \frac{1}{2}}\;t_f - {\textstyle \frac{3}{2}}.
\label{unilaw}\end{equation}
Thus the minimum energy emission occurs almost halfway to complete bundle 
collapse, in agreement with the simulations shown in Fig. \ref{En-uni} \\

\subsection{ Weibull distribution}

\begin{figure}
\includegraphics[width=6cm,height=6cm]{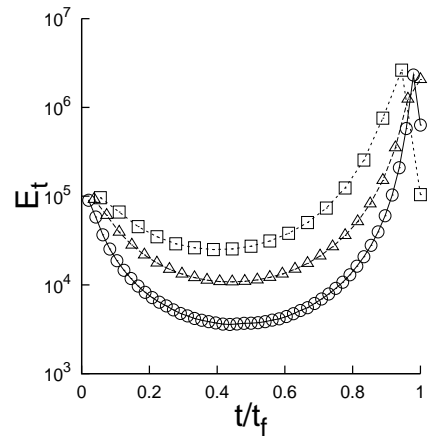}
\caption{The energy emission $E_t$ vs.\ the scaled step  
variable $t/t_f$ for a bundle of $N=10^7$ fibers obeying the Weibull threshold 
distribution (\ref{W}).
Different symbols are used for different excess stress levels
$\sigma -\sigma _c$: 0.001  (circles), 0.003 (triangles), 0.007 (squares).}
\label{En-Weibull}
\end{figure}

To illustrate the generality of the connection between the energy emission 
minimum and the bundle collapse, we now turn to a completely different 
threshold distribution, viz.\  a Weibull distribution of index 5,
\begin{equation}
P(x) = 1 - e^{-x^5}.
\label{W}\end{equation}
The critical stress for this distribution is $\sigma_c=(5e)^{-1/5}=0.5933994$.
Simulations reveal a qualitatively similar behavior as for the uniform 
distribution (Fig. \ref{En-Weibull} ).\\

The condition (\ref{min}) for minimal energy emission takes in this case the 
form
\begin{equation}
\ln n_t = 5^{5/2}\sigma^{35/2}\;n_t^{-20}\;\exp(-5\sigma^5/2n_t).
\label{Wcond} \end{equation}
It is straightforward to verify that for $\sigma=\sigma_c=(5e)^{-1/5}$ (\ref{Wcond}) has the solution $n_t=e^{-1/5}$. We are interested in a slightly overloaded bundle,
\begin{equation}
\sigma = \sigma_c+\epsilon= (5e)^{-1/5} +\epsilon,
\label{sigma}\end{equation}
with $\epsilon$ small and positive, and we seek the corresponding value of 
$n_t$. Putting 
(\ref{sigma}) and $n_t=e^{-1/5} +\delta$ into the condition (\ref{Wcond}), and 
expanding in $\epsilon$ and $\delta$, we obtain to first order in the small 
quantities
\begin{equation}
e^{1/5}\;\delta = 3(5e)^{1/5}\;\epsilon- {\textstyle \frac{7}{2}}\;e^{1/5}\;\delta.
\end{equation}
Thus $\delta = \frac{2}{3}\;5^{1/5}\;\epsilon$. Hence
\begin{equation}
n_t=e^{-1/5}\;+{\textstyle \frac{2}{3}}\;5^{1/5}\;\epsilon
\label{nW} \end{equation}
is the relative number of undamaged fibers when the energy emission is minimal.

To find $t_E$, the value of $t$ corresponding to (\ref{nW}), we take advantage of the ground work already done in \cite{Pradhan4} for the Weibull distribution. Eq.\ (29) in \cite{Pradhan4} shows that for small $\epsilon$ the iteration is of the form
\begin{equation}
n_t = e^{-1/5}\;-\;5^{1/5}\sqrt{\epsilon/C}\;\tan(t\sqrt{\epsilon C}-c),
\label{Wit}\end{equation}
with $C=\frac{5}{2}(5e)^{1/5}$ and the constant $c=\tan^{-1}[(1-e^{-1/5})\;5^{-1/5}\;\sqrt{C/\epsilon}]$ ensures that the initial condition $n_0=1$ is satisfied.

Comparison between (\ref{nW}) and (\ref{Wit}) gives
\begin{equation}
\tan(t_E\sqrt{\epsilon C}-c) = - {\textstyle \frac{2}{3}} \sqrt{\epsilon C}.
\end{equation}
To dominating order, then,
\begin{equation}
t_E\sqrt{\epsilon C} - c = - \tan^{-1}({\textstyle \frac{2}{3}} \sqrt{\epsilon C}) \simeq -{\textstyle \frac{2}{3}} \sqrt{\epsilon C}  
\end{equation}
For small $\epsilon$ the constant $c$ is very close to $\pi/2$, so that in 
good approximation we have
\begin{equation}
t_E = \frac{\pi}{2\sqrt{\epsilon C}}\;-\;{\textstyle \frac{2}{3}}.
\end{equation}

The collapse time $t_f$ was also evaluated in \cite{Pradhan4} to be
\begin{equation}
t_f =\frac{\pi}{\sqrt{\epsilon C}}.
\end{equation}
Consequently we have
\begin{equation}
t_E = {\textstyle \frac{1}{2}}\;t_f-{\textstyle \frac{2}{3}}.
\label{Wlaw}\end{equation}
To excellent approximation the minimum of energy emission occur halfway to the 
final collapse also for the Weibull distribution. There is every reason to 
believe that this feature is general.  
Note that since $t_E$ and $t_f$ are large numbers, the constants $\frac{3}{2}$ and $\frac{2}{3}$ in (\ref{unilaw}) and (\ref{Wlaw}) are of no significance.\\

\section{CONCLUDING REMARKS}
In summary, we have considered energy emission burst from slightly overloaded 
fiber bundles. During the degradation process there is a stage $t_E$ at which 
the energy emission is minimal, and we have demonstrated that the total bundle 
collapse occurs near $2t_E$. The demonstration has been performed merely for 
two very different distributions of the fiber breaking thresholds, but the 
result is doubtlessly universal. Thus the minimal energy emission gives an 
excellent estimate of when the bundle failure will take place. In our earlier 
work \cite{Pradhan6} we found that the fiber breaking rate has a minimum at 
half way to complete collapse. However, for practical purposes energy emission 
burst is a better entity to measure than the fiber breaking rate.

\begin{center}
{\bf Acknowledgement}
\end{center}

S. P. acknowledges financial support from Research Council of Norway
(NFR) through project number 199970/S60.

\begin{center}
{\bf APPENDIX}
\end{center}

In the main text we have considered  energy emissions from a slightly 
overloaded bundle, for which the breakdown process proceeds slowly. Here we 
give a brief description of where the energy emission minimum is located for 
larger loads. As a specific example we consider the model with  a uniform 
threshold distribution.

For a given stress $\sigma$ the relative number of unbroken fibers at the emission minimum is given by Eq.(18),
\begin{equation}
(1-n_t)n_t^2 = \sigma^{3/2}.
\label{A1}\end{equation}
Since the left-hand side of (\ref{A1}) does not exceed $4/27$, the maximum stress when an emission minimum is present, equals
\begin{equation}
\sigma_m = 2^{4/3}/9=0.279982.
\end{equation}
For a given stress, $\sigma_c\leq \sigma\leq \sigma_m$, Eq.(\ref{A1}) 
determines $n_t$, and the corresponding value of $t_E/t_f$ can be calculated 
using
\begin{equation}
n_t={\textstyle \frac{1}{2}}+\sqrt{\epsilon}\tan[B(1-2t_E/t_f)],
\end{equation}
with $B=\tan^{-1}(1/2\sqrt{\epsilon})$. We have combined Eqs.(21) and (23) in 
the main text. Fig.\ 3 shows the result. The position of the emission minimum, 
relative to $t_f$, decreases slowly with increasing stress
\begin{figure}[h]
\includegraphics[width=6cm,height=5cm]{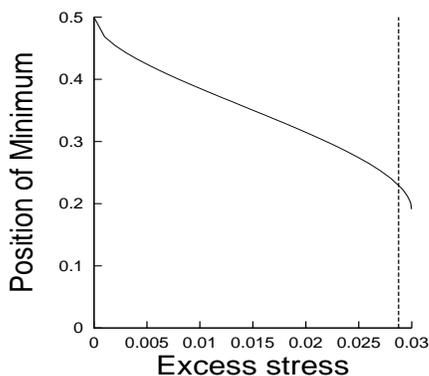}
\caption{Position of the emission minimum in terms of $t_E/t_f$ vs. excess 
stress for the uniform fiber strength distribution, computed using the
continuous time approximation. The dashed line marks the maximum stress with 
an emission minimum.} 
\label{min-position}
\end{figure}
In this derivation the integer $t$ is considered as a continuous variable. This
 treatment is  satisfactory  for slightly overloaded bundles, but less so when 
the $t_E$ is smaller. To find the smallest stress that produces a minimum, we 
consider the first few energy emissions. Eq.\ (8) for the uniform threshold 
distribution,
\begin{equation}
E_t = {\textstyle \frac{1}{6}}N\sigma^3 \left(n^{-3}_t-n^{-3}_{t-1}\right),
\end{equation}   
and the iteration (6), $n_0=1, n_1=1-\sigma, n_2=\frac{1-2\sigma}{1-\sigma}, \mbox{ etc.}$,
gives
\begin{eqnarray} 
E_0&=& \sigma^3\\
E_1&=& \left(\frac{\sigma}{1-\sigma}\right)^3-\sigma^3\\
E_2&= & \left(\frac{\sigma(1-\sigma)}{1-2\sigma}\right)^3- \left(\frac{\sigma}{1-\sigma}\right)^3\\
E_3& =& \left(\frac{\sigma(1-2\sigma)}{1-3\sigma+\sigma^2}\right)^3-\left(\frac{\sigma(1-\sigma)}{1-2\sigma}\right)^3,
\end{eqnarray}
where the common factor $N/6$ is omitted. For all subcritical stresses 
$E_1>E_0$, but a local minimum may occur already at $t=2$. It is easy to show 
that
\begin{eqnarray}
E_1&=&E_2 \mbox{ for } \sigma=\sigma_1=0.278764\\
 E_2&=&E_3 \mbox{ for } \sigma=\sigma_2=0.273054.
\end{eqnarray}
For $\sigma \geq \sigma_1$ the energy emission bursts increase monotonically 
in size. For $\sigma_2<\sigma<\sigma_1$ we have $E_2<E_1$ and also $E_2<E_3$.
Hence there is a local emission minimum at $t_E = 2$ in this range. 

Thus a local emission minimum is no longer present when the  stress value 
exceeds $\sigma_1$. 
For the limiting value $\sigma = \sigma_1$, the bundle collapse occurs
at $t_f=8$, signified by the first non-positive 
value of $n_t$. This gives $t_E/t_f=0.25$ at $\sigma =  \sigma_1$.
 
This limiting value is a little lower than the value $\sigma_m$, the end point 
of the graph in Fig. 3, obtained by the
 continuous procedure above. However, at the maximum stress $\sigma_1$ that 
produces a local emission minimum, the value of $t_E/t_f$  equals $0.230$, 
not far from the exact value
$0.25$. We conclude that the graph in Fig.\ 3 is  very precise for small 
supercritical stresses and fairly accurate up to the stress $\sigma_1$.


\begin{thebibliography}{99}
\bibitem{AE1} A. Petri, G. Paparo, A. Vespignani, A. Alippi and M. Costantini, Phys. Rev. Lett. {\bf 73} 3423 (1994). 
\bibitem{AE2} A. Garcimartin, A. Guarino, L. Bellon and S. Ciliberto, Phys. Rev. Lett. {\bf 79} 3202 (1997). 
\bibitem{Pradhan1} S.\ Pradhan, A.\ Hansen, and B.\ K.\ Chakrabarti, Rev.\ Mod.\ Phys. {\bf 82}, 499 (2010).
\bibitem{Herrmann} {\em Statistical models for the fracture of disordered materials}, edited by H.\ J.\ Herrmann ans S.\ Roux (Elsvier, Amsterdam, 1990).
\bibitem{Chakrabarti} B.\ K.\ Chakrabarti and L.\ G.\ Benguigui, {\em Statistical physics and  breakdown in disordered materials} (Oxford University Press, Oxford, 1997).
\bibitem{Sornette} D.\ Sornette, {\em Critical phenomena in natural sciences} (Springer-Verlag, Berlin, 2000)
\bibitem{Sahimi} M.\ Sahimi, {\em Heterogeneous materials II: Nonlinear and breakdown properties} (Springer-Verlag, Berlin 2003)
\bibitem{Bhatta} {\em Modeling critical and catastrophic phenomena in geoscience}, edited by P.\ Bhattacharyya and B.\ K.\ Chakraberti (Springer-Verlag, Berlin, 2006).
\bibitem{Hemmer1} P.\ C.\ Hemmer and A.\ Hansen, ASME J.\ Appl.\ Mech.\ {\bf 59}, 909 (1992).
\bibitem{Pradhan2} S.\ Pradhan, A.\ Hansen, and P.\ C.\ Hemmer, Phys.\ Rev.\ Lett. {\bf 95}, 125501 (2005); Phys.\ Rev.\ E {\bf 74}, 016122 (2006). 
\bibitem{Hemmer2} P.\ C.\ Hemmer, A.\ Hansen, and S.\ Pradhan, "Rupture processes in fiber bundle models", pp.\ 27-55 in [6].
\bibitem{Pradhan3} S.\ Pradhan and B.\ K.\ Chakrabarti, Phys.\ Rev. E {\bf 65}, 016113 (2001); S.\ Pradhan, P.\ Bhattacharyya, and B.\ K.\ Chakrabarti, Phys.\ Rev.\ {\bf 66}, 016116 (2002); P.\ Bhattacharyya, S.\ Pradhan, and B.\ K.\ Chakrabarti, Phys.\ Rev.\ {\bf 67}. 046122 (2003).
\bibitem{Pierce} F.\ T.\ Pierce, J.\ Text.\ Ind.\ {\bf 17}, 355 (1926).
\bibitem{Daniels} H.\ E.\ Daniels, Proc.\ Roy.\ Soc.\ London {\bf A. 183}, 405 (1945).
\bibitem{Smith} R.\ L.\ Smith, Ann.\ Prob.\ {\bf 10}, 137 (1982).
\bibitem{Phoenix} S.\ L.\ Phoenix and R.\ L.\ Smith, Int.\ J.\ Solids Struct. {\bf 19}, 479 (1983). 
\bibitem{Pradhan5} S.\ Pradhan and P.\ C.\ Hemmer, Phys.\ Rev.\ E {\bf 77}, 031138 (2008).
\bibitem{Pradhan4} S.\ Pradhan and P.\ C.\ Hemmer, Phys.\ Rev.\ E {\bf 75}, 056112 (2007).
\bibitem{Pradhan6} S.\ Pradhan and P.\ C.\ Hemmer, Phys.\ Rev.\ E {\bf 79}, 041148 (2009).
\end{thebibliography}
\end{document}